\documentclass[referee]{raa}            

\usepackage{graphicx,times}             
\usepackage{natbib}
\usepackage{amssymb,amsmath}
\bibpunct{(}{)}{;}{a}{}{,}
\usepackage{CJKutf8}
\usepackage[pagebackref=true]{hyperref}
\usepackage{subcaption}

\begin{document}
\begin{CJK}{UTF8}{gbsn}

  \title{ LAMOST medium-resolution spectroscopic survey of Galactic Open Clusters (LAMOST-MRS-O): An overview of survey plan and preliminary results 
}

   \setcounter{page}{1}          

   \author{Xi Zhang 
      \inst{1,2,3,5}
   \and Chengzhi Liu 
      \inst{1}
   \and Jing Zhong 
      \inst{3}
   \and Li Chen 
      \inst{2,3}
   \and Ali Luo 
      \inst{4}
   \and Jianrong Shi 
      \inst{4}
    \and Chao Liu 
      \inst{4}
   \and Jianjun Chen 
      \inst{4}
   \and Haotong Zhang 
      \inst{4}
 \and Jinliang Hou 
      \inst{2,3}
 \and LAMOST MRS Collaboration
   }

   \institute{Changchun Observatory, National Astronomical Observatories, Chinese Academy of Sciences, Changchun 130117, China\\
        \and
            School of Astronomy and Space Science, University of Chinese Academy of Sciences, No. 19A, Yuquan Road, Beijing 100049, China\\
        \and
            Astrophysics Division, Shanghai Astronomical Observatory, Chinese Academy of Sciences, 80 Nandan Road, Shanghai 200030, China\\
        \and
            CAS Key Laboratory of Optical Astronomy, National Astronomical Observatories, Chinese Academy of Sciences, Beijing 100101, China\\
        \and
            Institute of Astronomy and Information, Dali University, Dali 671003, China\\
        {\it e-mail: jzhong@shao.ac.cn \& chenli@shao.ac.cn}
}

\abstract{ As part of the LAMOST medium-resolution spectroscopic survey, the LAMOST-MRS-O is a non-time domain survey that aims to perform medium-resolution spectral observations for member stars in the open cluster area. This survey plans to obtain the spectroscopic parameters such as radial velocity and metal abundances of member stars and provide data support for further study on the chemical and dynamical characteristics and evolution of open clusters in combination with Gaia data. We have completed the observations on 10 open cluster fields and obtained 235184 medium-resolution spectra of 133792 stars. Based on the data analyzed of LAMOST DR11V1.1, for some clusters of particular concern, it is found that the sampling ratio of members stars with Gmag $<$ 15 mag can reach 70\%, which indicates that the LAMOST-MRS-O has reached our initial design goal.
}

   \authorrunning{Zhang et al.}            
   \titlerunning{LAMOST medium-resolution spectral survey of Open clusters}  

   \maketitle

%
%
\section{Introduction}           
\label{sect:intro}
The single-stellar-population origin of open clusters makes its member stars sharing many similar physical characteristics (such as age, distance, velocity, abundance, etc.). The cluster members include stars at various stages of evolution, such as various types of variable stars, binary stars, and stars with special astrophysical significance, e.g., hot subdwarfs \citep{2019ApJ...881....7L}, blue straggler stars \citep {2023A&A...672A..81L,2024A&A...686A.215L}, red clump stars \citep{2001AJ....121..327B}, etc. Open star clusters have become an ideal test place for the evolution theory of stars and binaries and provide crucial constraints for the stellar formation and evolution models.

Due to the wide age distribution (throughout the formation history of the Galactic disk) and mass distribution (from hundreds to tens of thousands of solar masses) of the open clusters, the detailed study of the properties of large sample open cluster members, such as the ratio of binary stars and their mass ratio distribution \citep {2022ApJ...930...44L}, mass function \citep{2024ApJ...971...71J}, and mass segregation \citep{2007AJ....134.1368C}, will further increase our understanding of the formation and evolution of the stellar system \citep {2015A&A...584A..91K,2019A&A...629L...4A}.

Since the release of Gaia DR2 data in 2018 \citep{2018A&A...616A...1G}, high-precision astrometry and photometry data have significantly improved the reliability of the determination results of open cluster members, thus extensively promoting the open cluster study. In addition, spectroscopic observation can provide more physical information about member stars, such as radial velocity, atmospheric parameters, element abundance, rotation velocity etc., \citep{2020A&A...640A.127Z,2022A&A...668A...4F,2020AJ....159..199D}.This information is also essential to better study the nature of open clusters and their member stars.

The LAMOST low-resolution spectroscopic survey (LAMOST-LRS) was officially started in 2012. It has obtained more than 20 million spectra and has a significant sky coverage (especially in the region of the Galactic anticenter). It has carried out uniform sampling and observation of a large number of star targets on the Galactic disk, including many open star clusters. \citep {2020A&A...640A.127Z} obtained the mean values of radial velocity and metallicity of 295 open clusters using the LAMOST low-resolution spectra (DR5). Using these clusters as a tracer, combined with the Gaia DR2 data, the radial and vertical metallicity gradients of the Galactic disk and their changes with age were studied. Furthermore,\citep {2022A&A...668A...4F} used LAMOST low-resolution spectra (DR8) and Gaia data to provide a catalog containing property parameters of 386 cluster and studied the abundance distribution of the Galaxy and the dynamic properties of the Galactic disk.

The LAMOST medium-resolution spectroscopic survey (LAMOST-MRS), officially started in September 2018, has a spectral resolution of R $\sim$ 7500 and wavelength coverage of 4950 \AA - 5350 \AA ~ at the blue part and  6300 \AA - 6800 \AA ~ at the red part. The LAMOST medium-resolution stellar spectrum can obtain nearly 20 chemical elements, such as lithium, carbon, sodium, magnesium, silicon, calcium, iron, manganese, nickel, etc., and the radial velocity with accuracy reached to nearly 1 km s$^{-1}$ \citep {2020arXiv200507210L}. 

As one of the LAMOST-MRS survey projects, LAMOST-MRS-O adopts multiple-visit observational modes with the same plate pointing but different fiber positions for the open cluster fields. This mode allows one to observe as many different cluster members as possible, improving the sampling rate of cluster member stars. LAMOST-MRS-O, obtaining the medium-resolution spectra of member stars, can provide radial velocity, atmospheric parameters, and element abundance more accurately than the low-resolution LAMOST spectra observed before. Furthermore, the LAMOST-MRS-O project aims to obtain a sample of cluster member spectra (Gmag = 9-15mag) with higher completeness, which will provide very important data support for studying the cluster properties and the stellar formation and evolution.

This paper mainly introduces the LAMOST-MRS-O spectroscopic survey project, including its scientific objectives, observing property parameters, and observed data qualities. 

\section{LAMOST-MRS-O survey}
LAMOST can simultaneously obtain spectral information of about 4000 sources and has unique advantages as large samples of spectroscopic surveys. However, for the specific observation of the open cluster field, according to the analysis of the results of LAMOST low-resolution survey data (DR5), the average sampling density in the low Galactic Latitude ($b$) region(-10$^\circ$ $< b < $ 10$^\circ$) is about 280 stars / deg$^2$. Due to the nearly uniform optical fiber distribution in the LAMOST focal plate, the sampled number of stars within the half number radius of most open clusters (with an average scale of about ten arc-minutes) is only about ten \citep {2020A&A...640A.127Z}. The low sampling rate of member stars will seriously restrict the relevant cluster research work. 

To solve the above problems, in September 2018, the LAMOST-MRS launched a sub-project of the open cluster observation (LAMOST-MRS-O). The observation magnitude range is from Gmag=9 - 15mag, and the sampling density of member stars can reach about 1000 - 2000 / deg$^2$. Considering that the optical fiber density of LAMOST is 200 / deg$^2$, it is planned to adopt the multiple visit mode for different member stars in the same cluster field. For each field, the total number of visits will be no less than eight, so the total number of observed member stars in the cluster area (about ten arc-minutes) is expected to be over 50. Based on this observation scheme, the statistical analysis of the cluster properties will be significantly improved. At the same time, under the LAMOST multiple-visit observation mode, some stars inevitably undergo repeated observations in the star cluster field. This will also provide valuable data for studying the time-domain properties of cluster members and field stars.

Compared with the LAMOST low-resolution spectrum, the LAMOST medium-resolution spectrum is expected to obtain more element abundances and stellar atmospheric parameters with higher accuracy. For example, under the condition of S/N greater than 10, the precision of radial velocity is about 1 km s$^{-1}$, [Fe/H] about 0.06 dex, other element abundance about 0.06 - 0.12 dex, stellar surface gravity about 0.17 dex and effective temperature about 110 K \citep{2020ApJ...891...23W}.

\subsection{Scientific goals}
In recent years, more and more open clusters have been discovered by using high-precision Gaia astrometry and photometric data \citep {2019AJ....157...12B,2020A&A...640A...1C, 2019ApJS..245...32L,2021A&A...652A.162C,2021RAA....21...45Q,2023ApJS..265...12Q, 
2021RAA....21...93H,2022ApJS..260....8H,2023A&A...673A.114H,2024A&A...686A..42H}. The number of star clusters has been greatly expanded. As the reliability and completeness of identifying open cluster members improves, an increasing number of clusters have been observed to exhibit tidal tail structures\citep {2019A&A...621L...2R,2019A&A...627A...4R, 2019A&A...627A.119C, 2020ApJ...889...99Z,2022RAA....22e5022B} and even the original extended structure \citep {2019A&A...624A..34Z}. In particular, in addition to having a much-extended region than before, it was found that the radial density distribution of a large portion of open clusters can be represented by a two-component distribution, which indicates that the extended structure of open clusters is not just a simple extension of the core components, but may have different origin or dynamical evolution characteristics. \citep {2022AJ....164...54Z} . These new understandings and discoveries of star clusters require more observations and in-depth research.

Combined with Gaia data, the LAMOST-MRS-O project will observe more cluster member spectra and obtain the radial velocity and atmospheric parameters of member stars, such as temperature, surface gravity, chemical abundance, etc., which will provide a unique contribution to a comprehensive understanding on physical properties of cluster member stars. Furthermore, based on the three-dimensional velocity, spatial distribution, and chemical abundance of member stars, one can also statistically study the dynamical disruption process\citep{2003AJ....126.2385O}, the cluster extended structure\citep {2019A&A...624A..34Z,2022RAA....22e5022B}, and the chemical evolution characteristic of star clusters \citep {2020A&A...640A.127Z}.

In addition, the LAMOST-MRS-O project may provide highly complete cluster member spectra for some clusters, especially in the outer region, which is of great scientific significance for the statistical study of various types of stars, such as Cepheid variable stars \citep {2021AcA....71..205P,2022A&A...668A..13H,2022ApJ...938...33L}, Be stars \citep {2012AstL...38..428T,2015RAA....15.1325L}, blue straggler stars \citep {2018MNRAS.481..226S,2020AJ....159...59R,2021A&A...650A..67R,2023A&A...672A..81L,
2024A&A...686A.215L}, red clump stars \citep {2017ApJ...840...77C,2020MNRAS.496.4637C,
2021A&A...655A..23M} and T Tauri stars \citep{2005ApJS..160..401P,2018ApJ...859....1M}, etc. In particular, for some unique phenomena of stellar evolution in open clusters, such as lithium depletion \citep {2010A&A...510A..50S,2021A&A...653A..72R} (Soderblom et al., 2010), extended main-sequence turn-offs \citep {2019ApJ...876...65L,2019ApJ...876..113S,2023MNRAS.525.5880H}, etc., the measurement results of lithium abundance and stellar rotation velocity from the LAMOST medium-resolution spectra will provide significant data support for these studies.

\subsection{Observation plan and strategy}
\begin{figure*}
    \centering
    \includegraphics[width=\textwidth, angle=0]{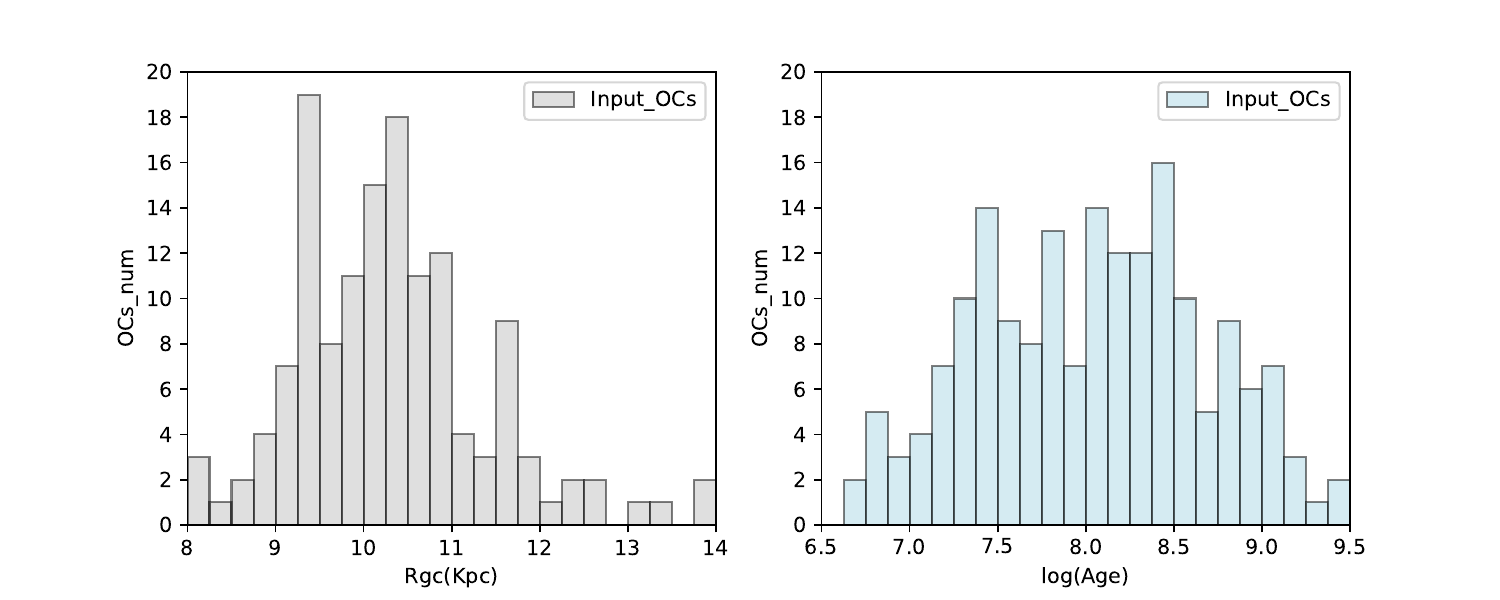}
    \caption{LAMOST-MRS-O Project plans to observe 18 cluster areas containing more than 180 clusters from CG18. Left panel: distribution of cluster distance from the Galactic center. Right panel: distribution of cluster ages in logarithm.}
    \label{ms2024-0394fig1}  
 \end{figure*}

In order to obtain a large number of spectra of cluster members in the LAMOST-MRS-O, we selected 18 open cluster related fields (hereafter OCs fields, as shown in Table ~\ref{ms2024-0394tab1}). Each OCs field contains at least 4-5 open clusters, with a total of about 180 clusters \citep {2018A&A...618A..93C} (hereafter CG18) and 397356 stars (Gmag$<$15 mag). Since most of OCs are located on the Galactic plane, our 18 OCs fields are arranged on the
low Galactic latitude,  mainly distributed in the direction of the Galactic-anti center (130$^\circ$ $< l <$ 240$^\circ$, and -10$^\circ$ $< b <$10$^\circ$, where $l,b$ represent the Galactic longitude and latitude respectively ), ranging from 8 to 14 kpc from the Galactic center, with ages between 30 Myr and 1 Gyr, as shown in Figure ~\ref{ms2024-0394fig1}. The overlap of each sky area shall not be greater than 10\%. Based on 20 square degrees of each sky area, the coverage area of the LAMOST-MRS-O is about 360 square degrees.

The LAMOST-MRS-O project was launched in September 2018 and ended in June 2023, lasting five years. The observed time in each year is from September of the current year to the middle of June of the following year. The LAMOST-MRS-O project is classified as a non-time domain (hereafter NT) observation mode, which is arranged on a gray night. Since an NT survey does not use single exposure information, we are able to co-add the single-exposure spectra for the same star to achieve higher S/N. Typically, a LAMOST-MRS-O plate will continuously take three single 1200s exposures and obtain the co-added spectra with a total exposure time of 3 × 1200s. This can obtain a limiting magnitude of G = 14.5 mag for the blue band and 15 mag for the red band with S/N～10. In terms of observation strategy, aiming at improving the sampling rate of member stars as much as possible, the completeness of obtained cluster members will increase to 70\% through 8 plates of spectroscopic observations, down to G=15mag.

\begin{table*}
\centering
\caption{The Plan of LAMOST-MRS-O Project Observation }\label{ms2024-0394tab1}
\begin{tabular}{cccccccc}
\hline\hline
Plan ID & Central Star&R.A.(2000)& Decl.(2000)& Nstar& Nocs & N\_memb &Remark\\
\hline
NT002740+583314C  & HIP 2191  & 00:27:40.51 & +58:33:14.08 & 23927 & 8  & 992&Y\\
NT014347+555239C  & HIP 8081  & 01:43:47.05 & +55:52:39.11 & 25538 & 4  & 885&Y\\ 
NT014355+602612C  & HD 10474  & 01:43:55.05 & +60:26:12.23 & 24349 & 13 & 1315& \\
NT021558+581737C  & HD 13744  & 02:15:58.69 & +58:17:37.00 & 24592 & 13 & 3186&Y\\
NT024142+552853C  & HIP 12575 & 02:41:42.76 & +55:28:53.76 & 18401 & 6  & 904&Y\\
NT024709+603414C  & HIP 13004 & 02:47:09.67 & +60:34:14.74 & 16096 & 20 & 1948& \\
NT033932+592643C  & HIP 17075 & 03:39:32.64 & +59:26:43.51 & 11761 & 5  & 782&Y\\
NT041712+505156C  & HIP 19986 & 04:17:12.30 & +50:51:56.95 & 13220 & 9  & 767&Y\\
NT043052+443610C  & HIP 21062 & 04:30:52.88 & +44:36:10.98 & 13944 & 4  & 923& \\
NT045451+440349C  & HIP 22842 & 04:54:51.22 & +44:03:39.56 & 18941 & 11 & 1446&Y\\
NT052826+344150C  & HIP 25624 & 05:28:26.39 & +34:41:50.93 & 20667 & 15 & 2961&Y\\
NT054955+314708C  & HIP 27538 & 05:49:55.53 & +31:47:08.66 & 23671 & 10 & 1616& \\
NT061047+133934C  & HIP 29310 & 06:10:47.36 & +13:39:34.09 & 20206 & 18 & 1227&Y\\
NT061156+231225C  & HIP 29425 & 06:11:56.25 & +23:12:25.42 & 20727 & 13 & 2178& \\
NT063435+045804C  & HIP 31363 & 06:34:35.61 & +04:58:04.87 & 21081 & 15 & 1801&Y\\
NT184324-054140C  & HD 173005 & 18:43:24.07 & -05:41:40.52 & 43882 & 19 & 3379& \\
NT194619+222809C  & HIP 97289 & 19:46:19.48 & +22:28:09.95 & 22572 & 11 & 1209& \\
NT200015+295514C  & HIP 98460 & 20:00:15.53 & +29:55:14.30 & 33955 & 9  & 1311& \\
\hline\hline
\multicolumn{8}{l}{{\sc Notes:}  This table provides details about the observation plan:}\\
\multicolumn{8}{l}{  "Nstar" refers to the number of all stars, while "Nmemb" represents the number of member stars in the}\\
\multicolumn{8}{l}{  open cluster that are planned for observation； }\\
\multicolumn{8}{l}{ The letter "Y" represents the area that has completed its observations up to now. }\\
\end{tabular}	
\end{table*}

\section{Survey data}
LAMOST DR11 v1.1 was released in September 2024 \footnote{http://www.lamost.org/dr11/v1.1/}, which included medium-resolution spectroscopy data from late October 2018 to June 2023. The  LAMOST DR11v1.1 provided over 10 million medium-resolution spectra, of which over 2.58 million spectra have atmospheric parameters. There are two kinds of stellar catalog present: the general catalog, which only includes general information including radial velocity parameter; the stellar parameter catalog, which additionally provides the atmospheric parameters and abundance of 13 different chemical elements.

In LAMOST DR11, ten OCs fields within LAMOST-MRS-O project have finished the observational task. Specifically, each of these OCs fields has been observed with eight different spectroscopic plates. The general catalog encompasses 235,184 combined spectra of 133,792 stars distributed in ten cluster areas, with the spectra incorporating details about either the blue end or the red end, while the S/N for all spectra are greater than 5. Furthermore, the stellar parameter catalog contains 105,068 spectra of 67,318 stars in the same cluster areas. Nearly all spectra have S/N greater than 10, providing radial velocity, atmospheric parameters, metallicity, and abundances of 13 other chemical elements, which greatly aids in-depth studies of cluster features and stellar evolution.

\section{Completion of observation}
\begin{figure*}
    \centering
   \includegraphics[width=\textwidth, angle=0]{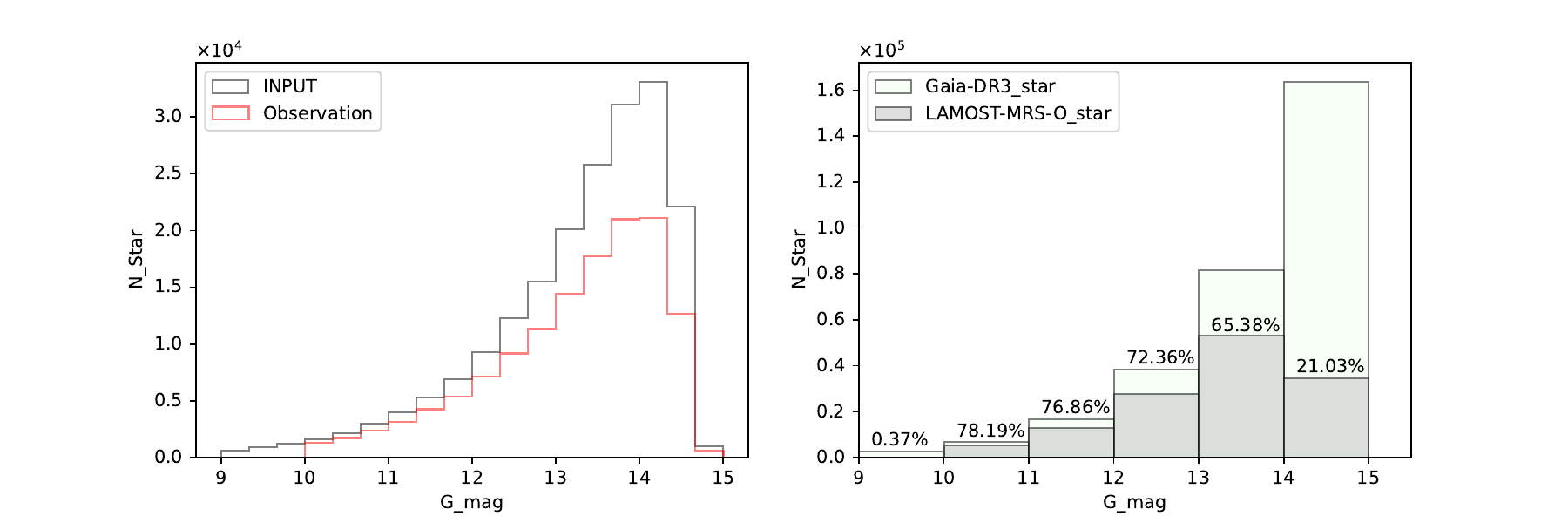} 
    \caption{The left panel presents histograms of input stars (in gray) and observed stars (in red), wherein the observation rate is around 70\% within the Gmag range spanning from 9 to 15. The right panel shows the histogram of observed stars and their completeness rate within each magnitude bin in comparison to Gaia DR3. For stars with a magnitude ranging from 10 to 13 mag, the completeness rate of the observed stars can exceed 75\%. }
   \label{ms2024-0394fig2}  
 \end{figure*}

\begin{figure*}
    \centering
    \includegraphics[scale=0.7]{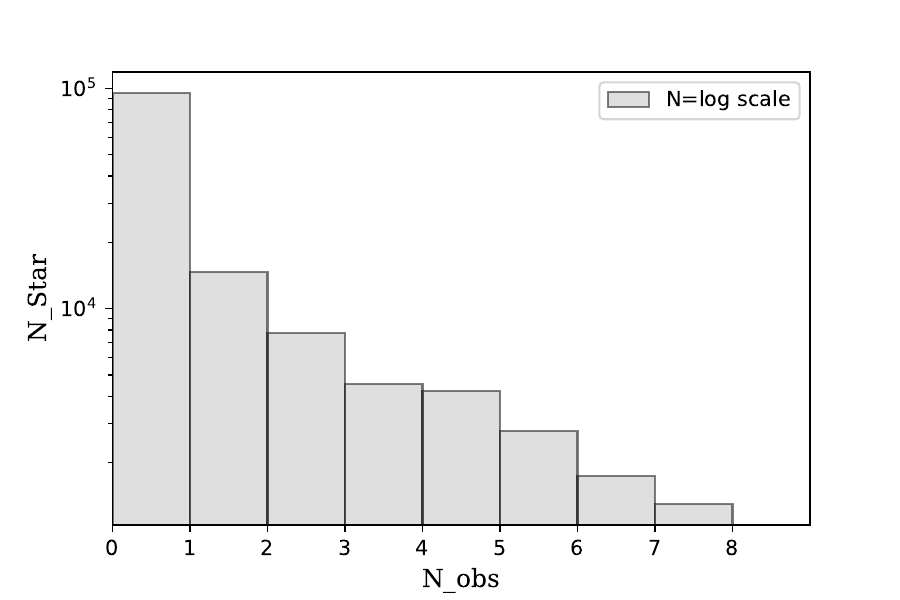}
    \caption{Distribution of the observation times of each star. The histogram indicates that the majority of stars were observed one or two times, and a small number of them were observed eight times, which is in line with the observational expectations.}
    \label{ms2024-0394fig3}  
 \end{figure*}

Within the magnitude range of Gmag = 9 - 15 mag, a majority of the stars in the input catalog have been successfully observed (refer to the left side of Figure ~\ref{ms2024-0394fig2}). This indicates that our observation plan has been effectively implemented. In the ten OCs fields, the scheduled number of stars for observation was 197,357, while the actual number of observed stars is 133,792. The completion rate has thus reached approximately 70\%. Figure ~\ref{ms2024-0394fig2} right panel presents the histogram of the observed stars and their corresponding completeness rates within each magnitude bin in comparison to Gaia DR3. It is evident that for stars possessing a magnitude ranging from 10 to 13 mag, the completeness rate of the observed stars can exceed 75\%. Furthermore, In Figure ~\ref{ms2024-0394fig3}, a histogram depicting the number of star observations in ten OCs fields is plotted. The majority of the stars within these areas have been observed at least 1 - 2 times, which is in accordance with the predefined observation strategy for the cluster areas. Regarding those stars that have been observed multiple times, their spectra can be utilized to investigate the variation of radial velocity as well as the stability of parameter measurement, thereby providing valuable insights for further astronomical research.

\begin{figure*}
    \centering
    \includegraphics[width=\textwidth, angle=0]{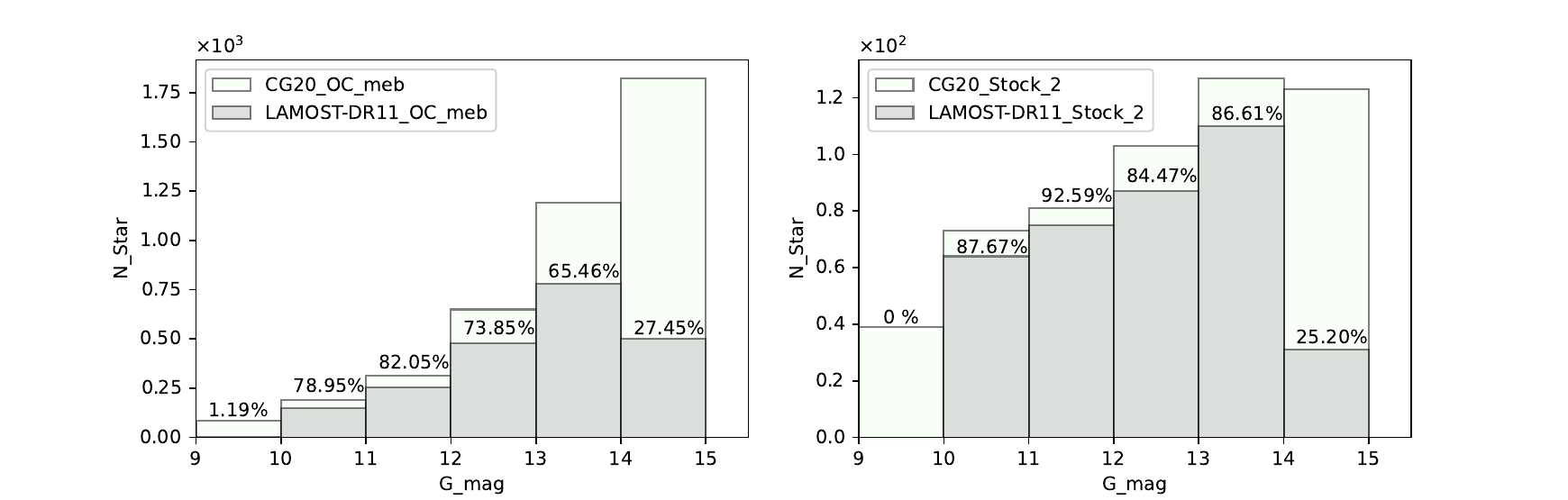}
    \caption{Distribution of the completion rate along the magnitude. The left panel presents the histogram of the observed cluster members, together with their corresponding completion rate in comparison to that of the CG20 members within the same regions.  The right panel  illustrates the observation completeness of Stock 2. Notably, it reveals that the observation completeness rate for stars with a magnitude brighter than 14 is greater than 80\%.
 }
    \label{ms2024-0394fig4}  
 \end{figure*}

To further investigate the observation completion rate of cluster members, we utilized the open cluster member catalog in \citep {2020A&A...640A...1C} (hereafter referred to as CG20). A cross-matching process was carried out with the LAMOST-MRS-O output catalog using a radius of 3 arcsec. As a result, we obtained 2,170 member stars belonging to 89 clusters. Among them, approximately 73\% of the member stars, which is about 1,577 in number, possess radial velocities measured by LAMOST. Figure~\ref{ms2024-0394fig4} shows the histogram of the observed cluster members and their completion rate in comparison with the CG20 members in the same areas. For most clusters in this region, about 70\% of the member stars with magnitudes between 10 and 13 mag have been observed.

As an example, we present the observation completeness of Stock 2 in the right panel of Figure~\ref{ms2024-0394fig4}. Within the magnitude range of G = 9 - 15 mag, there are 546 member stars in this cluster. Among them, 367 stars have been observed in LAMOST-MRS-O survey. Notably, the observation completeness rate for stars with magnitude less than 14 mag exceeds 80\%. Taking into account that the fiber distribution of LAMOST is near uniform, even with eight-time sampling conducted in the cluster-dense area, some member stars may still be missed. Nevertheless, for a cluster like stock 2, which has relatively scattered member stars, a relatively high sampling rate of its member stars can be achieved in the LAMOST-MRS-O observation mode.

\section{Parameters}
\subsection{Radial velocity}
The LAMOST MRS general catalog offers eight distinct kinds of radial velocity measurement results along with their corresponding measurement errors. We selected rv\_br1, which is obtained by combining the spectra from both the red and blue ends, as the radial velocity parameter for our sample stars. In the general catalog, a total of 235,184 combined spectra for 133,792 stars are included in 10 OCs fields. From this, we selected 97,479 stars with S/N $>$ 10 and rv\_br\_flag = 0. In this sample, approximately 25\% of the stars have been observed twice or more times, and their average radial velocity values will be weighted according to the error of each observation. For the radial velocity and its error of a single measurement, the data provided by the LAMOST MRS general star catalog are retained.  

\begin{figure*}
   \centering
   \includegraphics[scale=0.6]{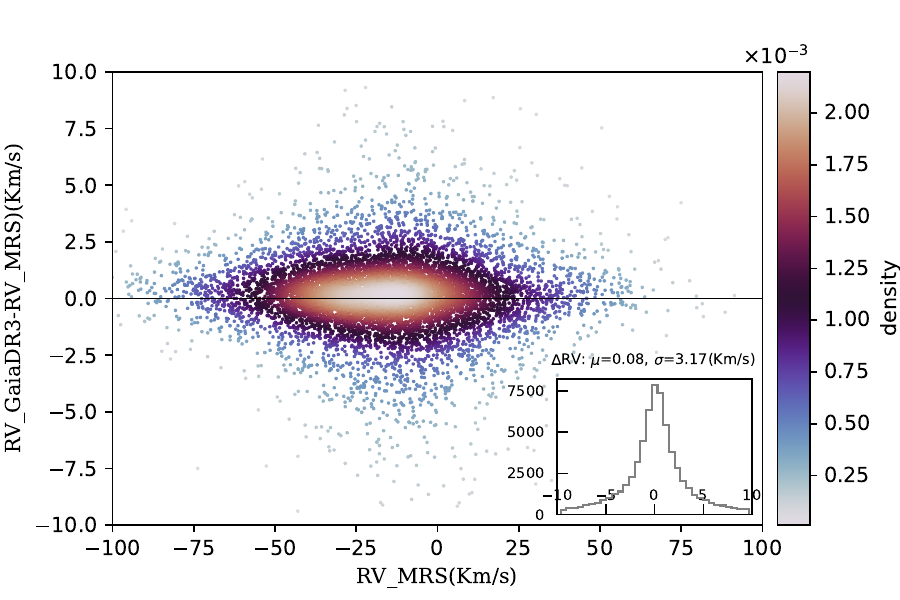}
   \caption{Difference in radial velocity between LAMOST-MRS and Gaia DR3. By calculating the radial velocity differences of 72,737 common stars, the difference value and standard deviation are only 0.08 km s$^{-1}$ and 3.09  s$^{-1}$ respectively.}
   \label{ms2024-0394fig5}  
\end{figure*}

We cross-matched the 97,479 stellar samples with the Gaia DR3 data and acquired the radial velocities of 72,737 common stars. As illustrated in Figure~\ref{ms2024-0394fig5}, the difference in radial velocity and its standard deviation between the LAMOST-MRS and Gaia DR3 are tiny, only 0.08 km s$^{-1}$ and 3.09 km s$^{-1}$ respectively. This indicates that the radial velocities of stars obtained from the LAMOST MRS spectra have high accuracy. 

\begin{figure*}
   \centering
   \includegraphics[width=\textwidth, angle=0]{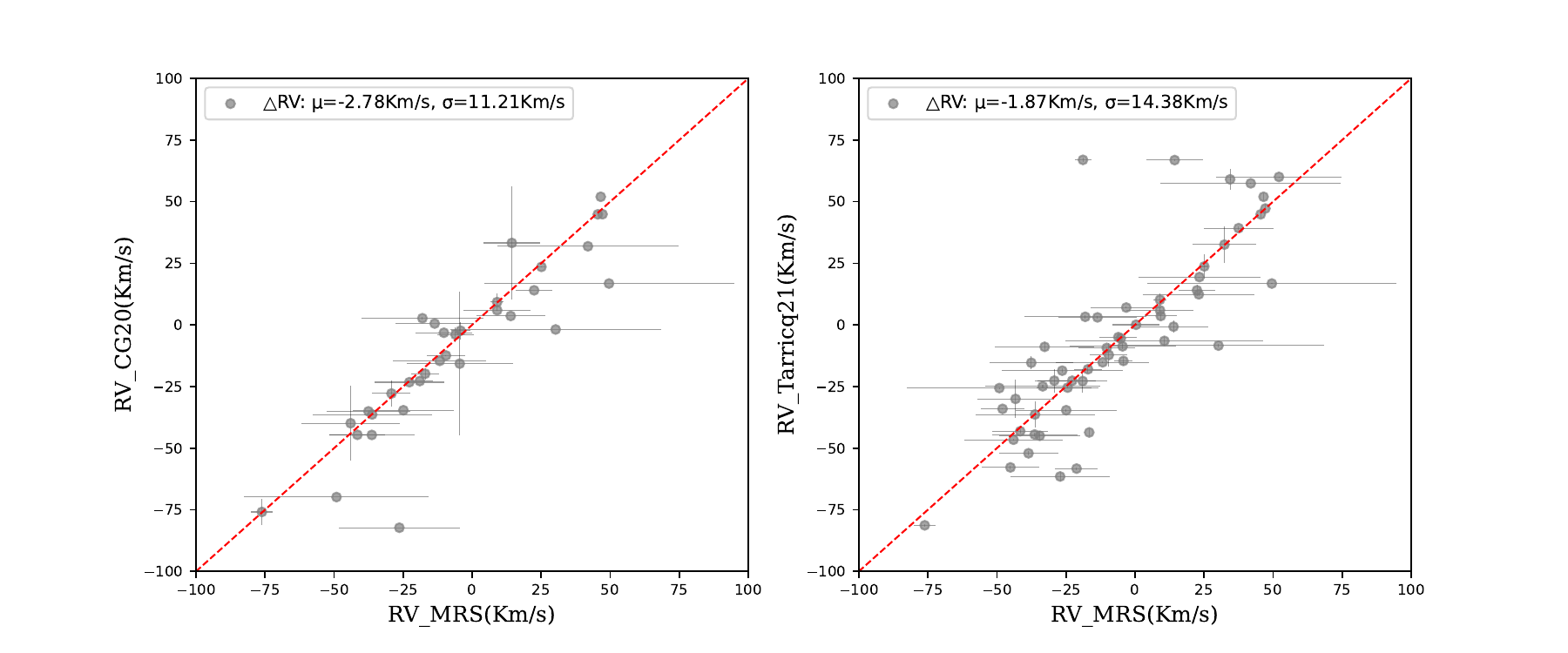}
   \caption{The left panel shows the comparison results of 33 common clusters. Their radial velocities, provided by CG20 and our sample, have small differences and dispersions, being -2.78 km s$^{-1}$ and 11.21 km s$^{-1}$ respectively. The right panel illustrates the differences and dispersions in radial velocities among 57 common clusters from \citet {2021A&A...647A..19T} and our samples, which are -1.87 km s$^{-1}$ and 14.38 km s$^{-1}$ respectively.}
   \label{ms2024-0394fig6}  
\end{figure*}

To inspect the average radial velocities of open star clusters, we cross-matched the stellar samples with the cluster member list of CG20. As a result, 1,577 member stars in 84 star clusters were obtained, among which 77 clusters possess two or more member stars. The radial velocity and its dispersion for each star cluster are represented by the mean and standard deviation of the radial velocity values of all member stars within the cluster. After excluding the clusters with remarkably large radial velocity dispersion (exceeding 50 km s$^{-1}$), our sample presently comprises 72 star clusters, among which 7 contain only one member star. Subsequently, we discovered that there are 33 common clusters having radial velocities provided by CG20. The left panel of Figure~\ref{ms2024-0394fig6} presents the comparison results of the two samples, showing small difference and dispersion in radial velocities, which are -2.78 km s$^{-1}$ and 11.21 km s$^{-1}$ respectively.

\citet {2021A&A...647A..19T} combined the spectroscopic data of Gaia DR2 \citep {2018A&A...616A...6S, 2019A&A...622A.205K, 2013Msngr.154...47R}, APOGEE DR16 \citep {2020ApJS..249....3A}, RAVE DR6 \citep {2020AJ....160...83S} and GALAH DR3 \citep {2020ApJ...888...75B}, and calculated the radial velocities of 1382 open clusters. After cross-matching with our sample, there are 57 common clusters, with radial velocity difference and dispersion of -1.87 km s$^{-1}$ and 14.38 km s$^{-1}$ respectively. Refer to the right panel in Figure~\ref{ms2024-0394fig6} for details.

\begin{figure*}
   \centering
   \includegraphics[width=\textwidth, angle=0]{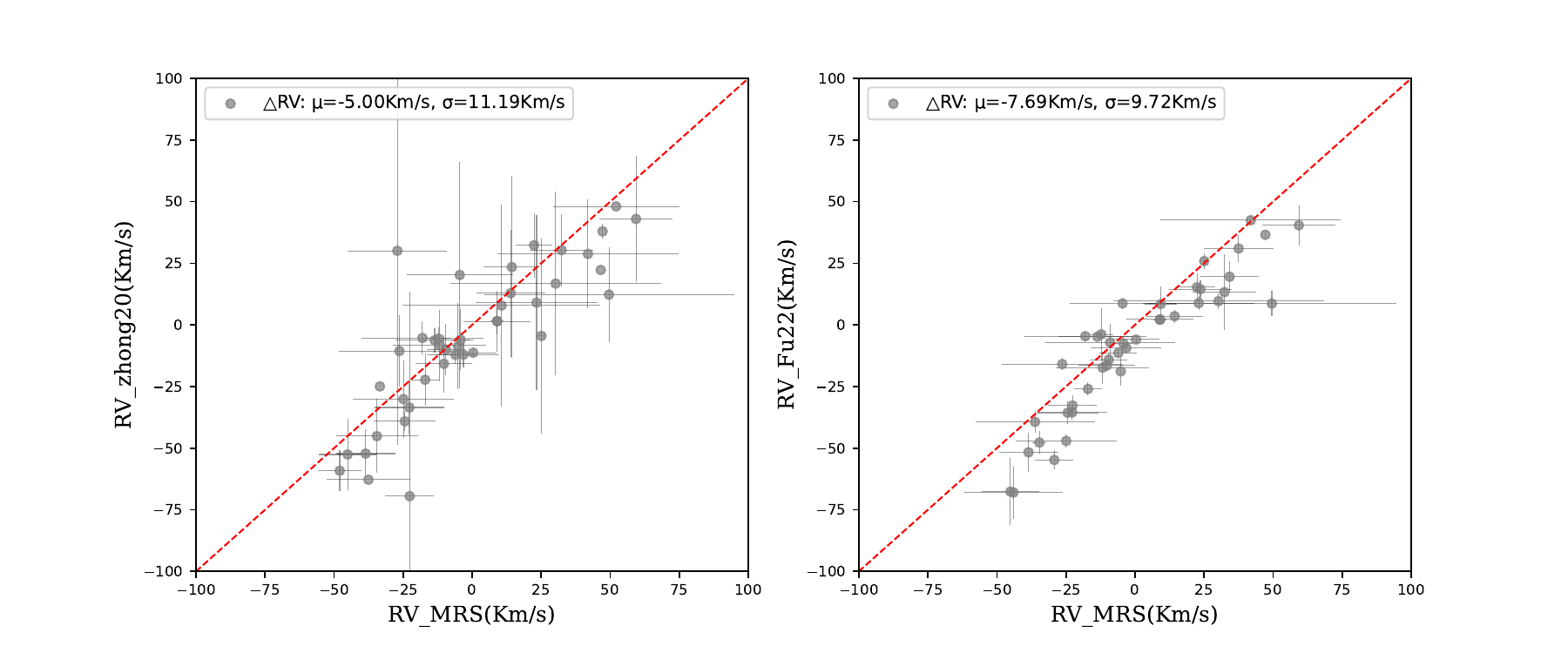}
   \caption{Comparison of radial velocity between the LAMOST low-resolution spectra results and our samples was conducted. Regarding the results from Zhong20 (with 42 common clusters) and Fu22 (with 41 common clusters), the difference results for comparison are -5.0 km s$^{-1}$ and -7.69 km s$^{-1}$, respectively, whereas the deviations are 11.19 km s$^{-1}$ and 9.72 km s$^{-1}$, respectively.}
   \label{ms2024-0394fig7}  
\end{figure*}

Additionally, by making use of the LAMOST DR5 and DR8 low-resolution spectra, \citep{2020A&A...640A.127Z} and \citep{2022A&A...668A...4F} obtained the average radial velocities of 295 and 386 clusters, respectively. After cross-matching our sample with the tables presented in the two papers, we found 42 and 41 common clusters, respectively. The comparison differences are -5.00 km s$^{-1}$ and -7.69 km s$^{-1}$, with deviations of 11.19 km s$^{-1}$ and 9.72 km s$^{-1}$ respectively, as illustrated in Figure ~\ref{ms2024-0394fig7}.

In general, compared with the recent literature results, the radial velocity difference and dispersion of stars acquired through LAMOST-MRS-O exhibit relatively small values. This is essentially consistent with the conclusion proposed by \citep {2019ApJS..244...27W}, showing an internal accuracy of 1.36  km s$^{-1}$ for S/N $\ge$ 10. It is worth noting that, in addition to the observational error, the relatively large average radial velocity dispersion within some open clusters might be attributed to the influence of binary members and the discrepant measurement errors of stars with different colors.

\subsection{Metallicity}
From the LAMOST-MRS DR11v1.1 stellar parameter catalogue, we acquired the [M/H] abundance as well as the abundance parameters of 13 chemical elements for 67,318 stars within the ten completed observational fields. We utilized this sample to conduct an analysis on the metal abundance of targeted LAMOST-MRS-O open star clusters. Given that the LAMOST DR11v1.1 stellar parameter catalogue fails to provide the observational errors of element abundances, for the stars that have been observed multiple times, their metal abundances are represented by the mean and standard deviation of the observed data.  For each cluster, the average metal abundance and its associated error are signified by the mean and standard deviation of the abundances of its member stars.

\begin{figure*}
   \centering
   \includegraphics[width=\textwidth, angle=0]{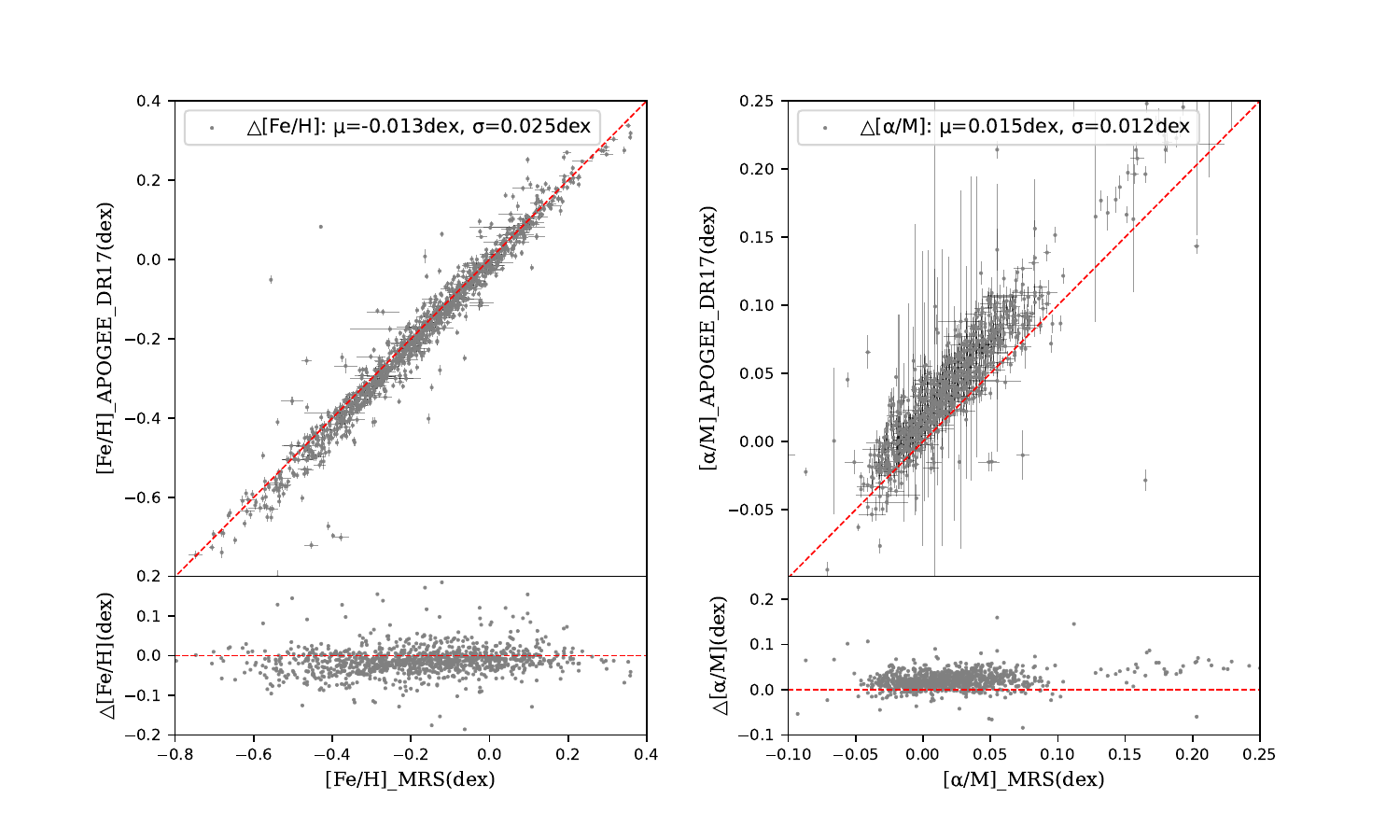}    
    \caption{ Comparison between APOGEE DR17 and our catalog. There are 1137 stars sharing [Fe/H] values and 968 stars sharing [$\alpha$/M] values in common. The left and right panels illustrate that the mean value of $\Delta$[Fe/H] is -0.013 dex and that of $\Delta$[$\alpha$/M] is 0.015 dex, with the standard deviations being 0.025 dex and 0.012 dex, respectively.}
 \label{ms2024-0394fig8}  
\end{figure*}

\begin{figure*}
     \centering
     \includegraphics[width=\textwidth, angle=0]{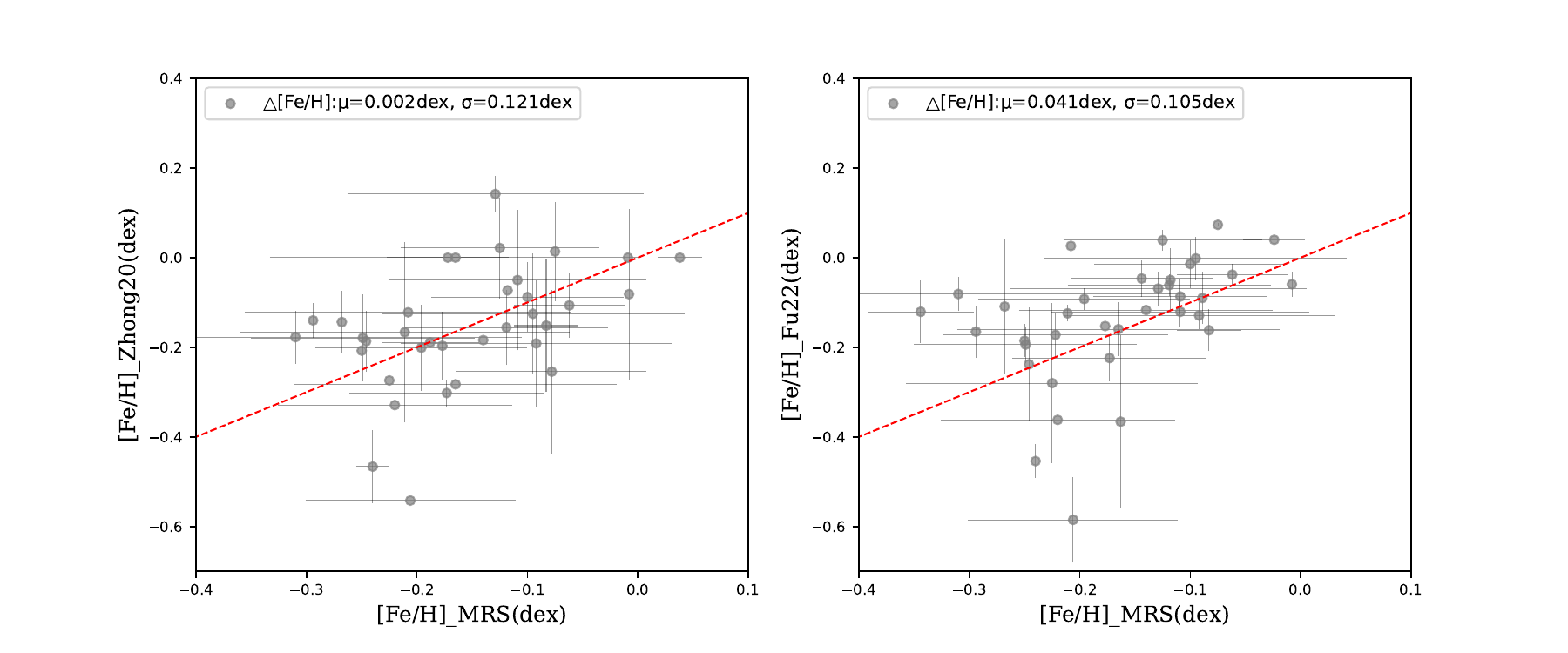}
     \caption{ Discrepancies in [Fe/H] of common clusters between the literature and our samples are presented as follows. In the left panel, when they are compared with \citep{2020A&A...640A.127Z}, which involves 36 common clusters, the mean value amounts to approximately 0.002 dexes, and the standard deviation equals approximately 0.121 dex. In the right panel, when they are contrasted with \citep{2022A&A...668A...4F}, which pertains to 37 common clusters, the mean value is about 0.041 dex and the standard deviation approximates to 0.105 dex.}
    \label{ms2024-0394fig9}  
\end{figure*}

\begin{table*}
\centering
\caption{Elemental abundances of Stock 2 from the LAMOST-MRS-O survey data}\label{ms2024-0394tab2}
\begin{tabular}{cccc}
\hline\hline
Element & $\mu$ (dex)& $\sigma$ (dex) & N\_member  \\
\hline
$[M/H]$           & -0.066    & 0.050    & 200  \\
$[\alpha / M]$    & -0.005    & 0.035    & 200  \\
$[Fe/H]$          & -0.061    & 0.051    & 200  \\
$[C/Fe]$          & -0.029    & 0.059    & 200  \\
$[N/Fe]$          & -0.002    & 0.071    & 200  \\
$[Mg/Fe]$         & -0.005    & 0.065    & 200  \\
$[Si/Fe]$         & -0.020    & 0.051    & 200  \\ 
$[Ca/Fe]$         &  0.012    & 0.044    & 200  \\ 
$[Ni/Fe]$         & -0.025    & 0.024    & 200  \\
$[O/Fe]$          &  0.043    & 0.010    & 2    \\
$[Al/Fe]$         & -0.017    & 0.029    & 2    \\
$[S/Fe]$          &  0.063    & 0.031    & 2    \\
$[Ti/Fe]$         & -0.049    & 0.027    & 2    \\
$[Cr/Fe]$         & -0.009    & 0.011    & 2    \\ 
$[Cu/Fe]$         &  0.015    & 0.019    & 2    \\ 
\hline\hline
\multicolumn{4}{l}{ }\\
\end{tabular}	
\end{table*}

The sample of 67,318 stars was cross-matched with the high-resolution APOGEE DR17, resulting in 1,137 stars with [Fe/H], 904 stars with [M/H] and 968 stars with [$\alpha$/M] values in common. The mean values of $\Delta$[Fe/H] and $\Delta$[$\alpha$/M] are -0.013 dex and 0.015 dex respectively, with corresponding standard deviations of 0.025 dex and 0.012 dex. As depicted in Figure ~\ref{ms2024-0394fig8}, the chemical abundance values provided by LAMOST-MRS are in accordance with the APOGEE results. 

We further cross-matched the sample of 67,318 stars with CG20 to obtain 555 member stars belonging to 62 open star clusters. These 62 clusters were cross-matched with the literature \citep {2020A&A...640A.127Z} and \citep {2022A&A...668A...4F} separately, yielding 36 and 37 common clusters with [Fe/H] parameter available, respectively. The comparison results of metallicity parameters in open star clusters are shown in Figure ~\ref{ms2024-0394fig9}. When compared with \citep{2020A&A...640A.127Z}, the mean value of $\Delta$[Fe/H] is approximately 0.002 dex and the standard derivation is around 0.121 dex, which is slightly lower than the corresponding mean difference value of approximately 0.041 dex and the standard deviation of around 0.105 dex when compared with \citep{2022A&A...668A...4F}. 

\begin{figure*}
     \centering
     \includegraphics[width=\textwidth, angle=0]{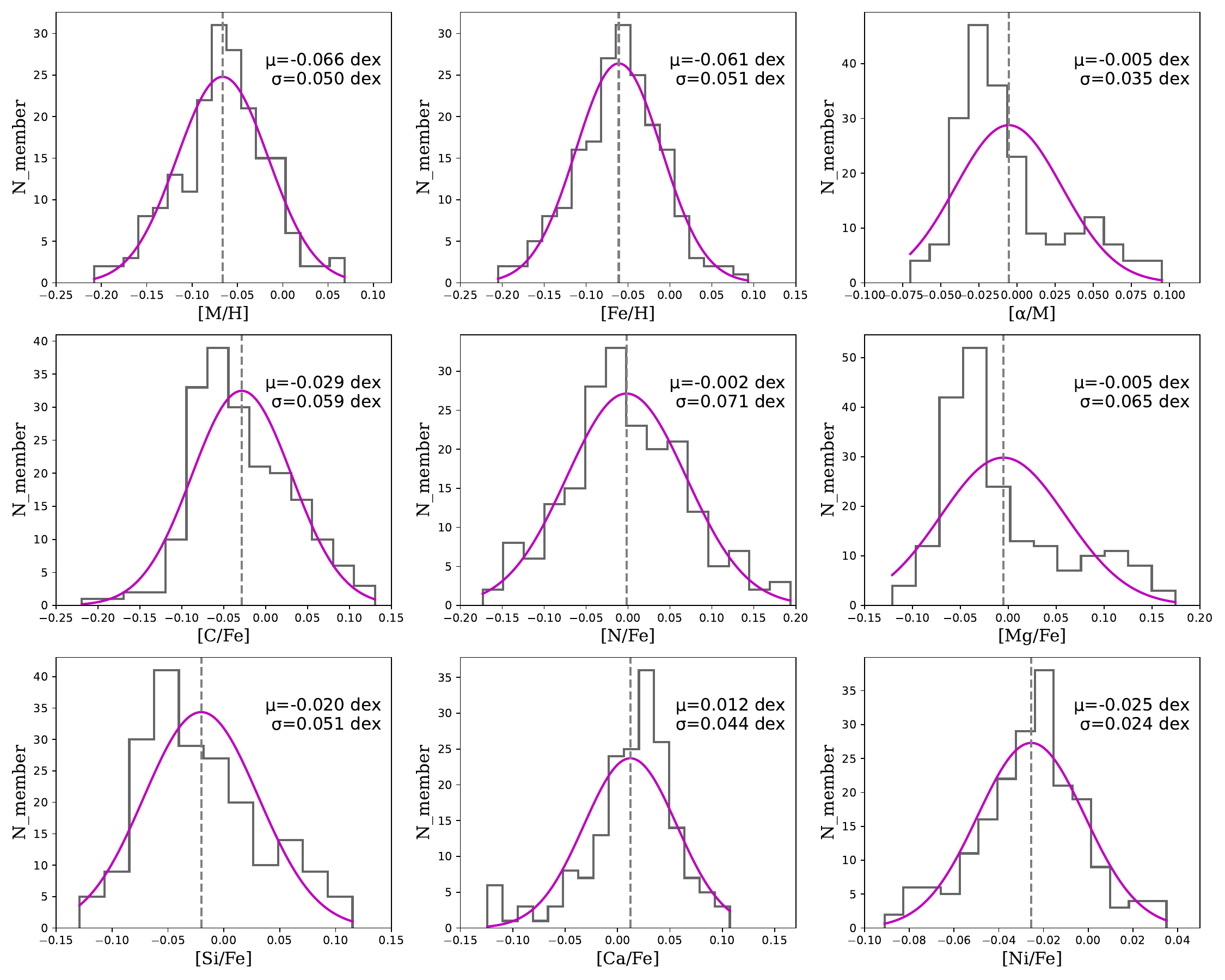}
     \caption{ Histograms and Gaussian profile for abundance parameters which have sufficient member spectra. In general, there is no significant bias and the LAMOST-MRS spectroscopic measurements are of high precision. }
    \label{ms2024-0394fig10}  
\end{figure*}

To examine the precision of the elemental abundance results of LAMOST-MRS, we selected a cluster Stock 2 to examplify the consistency of the elemental abundance. In the LAMOST-MRS-O survey, there are 304 member stars in Stock 2 that have spectral observations. Of these member stars, most of them provide elemental abundances, including iron, carbon, nitrogen, magnesium, silicon, calcium, nickel, etc. Table~\ref{ms2024-0394tab2} lists the mean value and standard deviation of these elements in Stock 2. It can be seen that for nine abundance parameters there are sufficient member stars in Stock 2 which have spectroscopic observations, while for other six abundances only two members have spectroscopic data available. In Figure~\ref{ms2024-0394fig10}, we plotted the metallicity distribution histograms and conducted Gaussian fittings for all the abundance parameters. Most abundance parameters display a good Gaussian distribution and a small standard deviation, indicating that there is no significant bias and the LAMOST-MRS spectroscopic measurements are of high precision. 

\begin{figure*}
     \centering
     \includegraphics[scale=0.7]{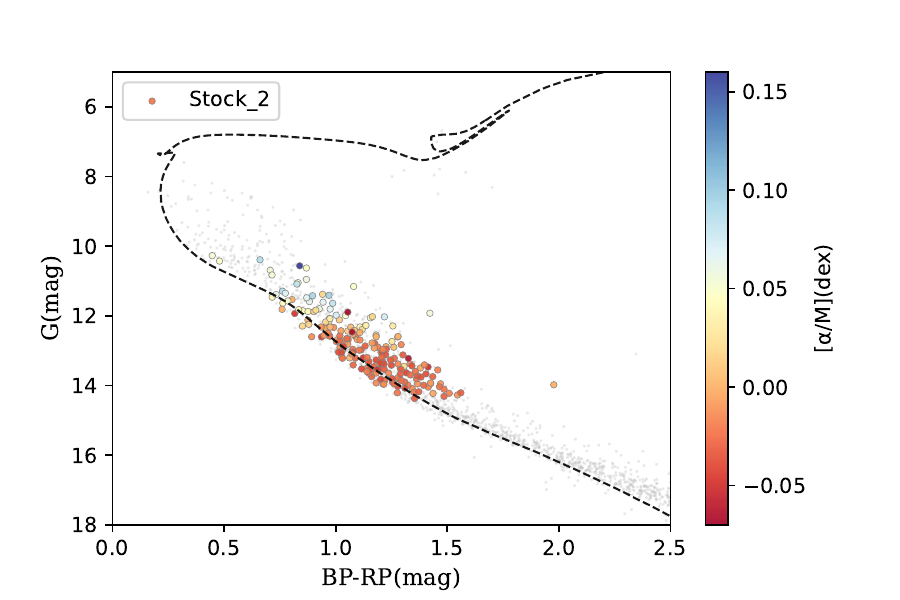}
     \caption{Spectroscopic member distribution in the color-magnitude diagram of Stock 2. Colors represent the [$\alpha$/M] abundance. Moreover, the [$\alpha$/M]  measurements of hot stars tend to display large deviations in the LAMOST-MRS catalog. }
    \label{ms2024-0394fig11}  
\end{figure*}

However, we observed that the distribution of the parameters exhibited an extended tail, particularly regarding the two parameters, [$\alpha$/M] and [Mg/Fe]. Figure~\ref{ms2024-0394fig11} shows the Stock 2 spectroscopic member distribution in the color-magnitude diagram, where colors represent the [$\alpha$/M] abundance. It is clear that the extended tail of [$\alpha$/M] is strongly correlated to the stellar type: the abundance measurements of hot stars tend to exhibit large deviations.

\section{Summary}
We introduced the LAMOST-MRS-O survey project, which encompasses the scientific objectives, the observational plan and strategy, the observational data, as well as the completion status of the observations. The high-completeness observation mode of LAMOST-MRS-O survey targeting over 100 Galactic open clusters is designed to acquire high-precision radial velocities, stellar parameters, and abundances of multiple elements for the cluster members. Through further integration of the distance and proper motion data from Gaia, high-precision three-dimensional spatial positions and three-dimensional velocities of the stars will be obtained, thereby significantly enhancing the reliability of cluster member determination. Additionally, by combining the multi-color photometric data of Gaia with the LAMOST spectral data of a large sample of cluster members, an effective and fully purified color-magnitude diagram can be constructed, enabling the derivation of accurate cluster ages, distances, mass distributions, and kinematic properties, which in turn provides comprehensive data support and observational constraints for stellar physics studies based on clusters.

\begin{acknowledgements}
This work is supported by the National Natural Science Foundation of China (NSFC) through grants 12090040, 12090042, 12073060, and the National Key R\&D Program of China No. 2019YFA0405501.
Jing Zhong acknowledges the Youth Innovation Promotion Association CAS, the Science and Technology Commission of Shanghai Municipality (Grant No. 22dz1202400), and the Program of Shanghai Academic/Technology Research Leader.
Li Chen acknowledges the science research grants from the China Manned Space Project with NO. CMS-CSST-2021-A08.
Guoshoujing Telescope (the Large Sky Area Multi-Object Fiber Spectroscopic Telescope LAMOST) is a National Major Scientific Project built by the Chinese Academy of Sciences. Funding for the project has been provided by the National  Development and Reform Commission. LAMOST is operated and managed by the National Astronomical Observatories, Chinese Academy of Sciences.
\end{acknowledgements}

\end{CJK}


\begin{thebibliography}{}

\bibitem[Ahumada et al.(2020)]{2020ApJS..249....3A} Ahumada, R., Allende Prieto, C., Almeida, A., et al.\ 2020, \apjs, 249, 3. doi:10.3847/1538-4365/ab929e

\bibitem[Andr{\'e} et al.(2019)]{2019A&A...629L...4A} Andr{\'e}, P., Arzoumanian, D., K{\"o}nyves, V., et al.\ 2019, \aap, 629, L4. doi:10.1051/0004-6361/201935915

\bibitem[Bai et al.(2022)]{2022RAA....22e5022B} Bai, L., Zhong, J., Chen, L., et al.\ 2022, Research in Astronomy and Astrophysics, 22, 055022. doi:10.1088/1674-4527/ac60d2

\bibitem[Barros et al.(2020)]{2020ApJ...888...75B} Barros, D.~A., P{\'e}rez-Villegas, A., L{\'e}pine, J.~R.~D., et al.\ 2020, \apj, 888, 75. doi:10.3847/1538-4357/ab59d1

\bibitem[Bica et al.(2019)]{2019AJ....157...12B} Bica, E., Pavani, D.~B., Bonatto, C.~J., et al.\ 2019, \aj, 157, 12. doi:10.3847/1538-3881/aaef8d

\bibitem[Bragaglia et al.(2001)]{2001AJ....121..327B} Bragaglia, A., Carretta, E., Gratton, R.~G., et al.\ 2001, \aj, 121, 327. doi:10.1086/318042


\bibitem[Cantat-Gaudin et al.(2018)]{2018A&A...618A..93C} Cantat-Gaudin, T., Jordi, C., Vallenari, A., et al.\ 2018, \aap, 618, A93. doi:10.1051/0004-6361/201833476

\bibitem[Cantat-Gaudin et al.(2020)] {2020A&A...640A...1C} Cantat-Gaudin, T., Anders, F., Castro-Ginard, A., et al.\ 2020, \aap, 640, A1. doi:10.1051/0004-6361/202038192

\bibitem[Carrera et al.(2019)]{2019A&A...627A.119C} Carrera, R., Pasquato, M., Vallenari, A., et al.\ 2019, \aap, 627, A119. doi:10.1051/0004-6361/201935599

\bibitem[Castro-Ginard et al.(2021)]{2021A&A...652A.162C} Castro-Ginard, A., McMillan, P.~J., Luri, X., et al.\ 2021, \aap, 652, A162. doi:10.1051/0004-6361/202039751

\bibitem[Chen et al.(2020)]{2020MNRAS.496.4637C} Chen, B., Wang, S., Hou, L., et al.\ 2020, \mnras, 496, 4637. doi:10.1093/mnras/staa1827

\bibitem[Chen et al.(2007)]{2007AJ....134.1368C} Chen, L., de Grijs, R., \& Zhao, J.~L.\ 2007, \aj, 134, 1368. doi:10.1086/521022

\bibitem[Chen et al.(2017)]{2017ApJ...840...77C} Chen, Y.~Q., Casagrande, L., Zhao, G., et al.\ 2017, \apj, 840, 77. doi:10.3847/1538-4357/aa6d0f

\bibitem[Donor et al.(2020)]{2020AJ....159..199D} Donor, J., Frinchaboy, P.~M., Cunha, K., et al.\ 2020, \aj, 159, 199. doi:10.3847/1538-3881/ab77bc

\bibitem[Fu et al.(2022)]{2022A&A...668A...4F} Fu, X., Bragaglia, A., Liu, C., et al.\ 2022, \aap, 668, A4. doi:10.1051/0004-6361/202243590

\bibitem[Gaia Collaboration et al.(2018)]{2018A&A...616A...1G} Gaia Collaboration, Brown, A.~G.~A., Vallenari, A., et al.\ 2018, \aap, 616, A1. doi:10.1051/0004-6361/201833051

\bibitem[Hao et al.(2022)]{2022A&A...668A..13H} Hao, C.~J., Xu, Y., Wu, Z.~Y., et al.\ 2022, \aap, 668, A13. doi:10.1051/0004-6361/202244570

\bibitem[He et al.(2023)]{2023MNRAS.525.5880H} He, C., Li, C., Sun, W., et al.\ 2023, \mnras, 525, 5880. doi:10.1093/mnras/stad2674

\bibitem[He et al.(2021)] {2021RAA....21...93H} He, Z.-H., Xu, Y., Hao, C.-J., et al.\ 2021, Research in Astronomy and Astrophysics, 21, 093. doi:10.1088/1674-4527/21/4/93

\bibitem[He et al.(2022)]{2022ApJS..260....8H} He, Z., Li, C., Zhong, J., et al.\ 2022, \apjs, 260, 8. doi:10.3847/1538-4365/ac5cbb

\bibitem[Hunt \& Reffert(2023)]{2023A&A...673A.114H} Hunt, E.~L. \& Reffert, S.\ 2023, \aap, 673, A114. doi:10.1051/0004-6361/202346285

\bibitem[Hunt \& Reffert(2024)]{2024A&A...686A..42H} Hunt, E.~L. \& Reffert, S.\ 2024, \aap, 686, A42. doi:10.1051/0004-6361/202348662

\bibitem[Jiang et al.(2024)]{2024ApJ...971...71J} Jiang, Y., Zhong, J., Qin, S., et al.\ 2024, \apj, 971, 71. doi:10.3847/1538-4357/ad5344


\bibitem[Katz et al.(2019)]{2019A&A...622A.205K} Katz, D., Sartoretti, P., Cropper, M., et al.\ 2019, \aap, 622, A205. doi:10.1051/0004-6361/201833273

\bibitem[K{\"o}nyves et al.(2015)]{2015A&A...584A..91K} K{\"o}nyves, V., Andr{\'e}, P., Men'shchikov, A., et al.\ 2015, \aap, 584, A91. doi:10.1051/0004-6361/201525861

\bibitem[Lin et al.(2015)]{2015RAA....15.1325L} Lin, C.-C., Hou, J.-L., Chen, L., et al.\ 2015, Research in Astronomy and Astrophysics, 15, 1325. doi:10.1088/1674-4527/15/8/015

\bibitem[Li et al.(2023)]{2023A&A...672A..81L} Li, C., Zhong, J., Qin, S., et al.\ 2023, \aap, 672, A81. doi:10.1051/0004-6361/202244998

\bibitem[Li et al.(2024)]{2024A&A...686A.215L} Li, C., Zhong, J., Qin, S., et al.\ 2024, \aap, 686, 
A215. doi:10.1051/0004-6361/202449393

\bibitem[Li et al.(2019)]{2019ApJ...876...65L} Li, C., Sun, W., de Grijs, R., et al.\ 2019, \apj, 876, 65. doi:10.3847/1538-4357/ab15d2

\bibitem[Li \& Shao(2022)]{2022ApJ...930...44L} Li, L. \& Shao, Z.\ 2022, \apj, 930, 44. doi:10.3847/1538-4357/ac5f4f

\bibitem[lin et al.(2022)]{2022ApJ...938...33L} lin, Z., Xu, Y., Hao, C., et al.\ 2022, \apj, 938, 33. doi:10.3847/1538-4357/ac9051

\bibitem[Liu\& Pang(2019)] {2019ApJS..245...32L} Liu, L.\& Pang, X.\ 2019, \apjs, 245, 32. doi:10.3847/1538-4365/ab530a

\bibitem[Liu et al.(2020)]{2020arXiv200507210L} Liu, C., Fu, J., Shi, J., et al.\ 2020, arXiv:2005.07210. doi:10.48550/arXiv.2005.07210

\bibitem[Luo et al.(2019)]{2019ApJ...881....7L} Luo, Y., N{\'e}meth, P., Deng, L., et al.\ 2019, \apj, 881, 7. doi:10.3847/1538-4357/ab298d

\bibitem[Magrini et al.(2021)]{2021A&A...655A..23M} Magrini, L., Smiljanic, R., Franciosini, E., et al.\ 2021, \aap, 655, A23. doi:10.1051/0004-6361/202141275

\bibitem[Mauc{\'o} et al.(2018)]{2018ApJ...859....1M} Mauc{\'o}, K., Brice{\~n}o, C., Calvet, N., et al.\ 2018, \apj, 859, 1. doi:10.3847/1538-4357/aabf40

\bibitem[Odenkirchen et al.(2003)]{2003AJ....126.2385O} Odenkirchen, M., Grebel, E.~K., Dehnen, W., et al.\ 2003, \aj, 126, 2385. doi:10.1086/378601


\bibitem[Pietrukowicz et al.(2021)]{2021AcA....71..205P} Pietrukowicz, P., Soszy{\'n}ski, I., \& Udalski, A.\ 2021, \actaa, 71, 205. doi:10.32023/0001-5237/71.3.2

\bibitem[Preibisch et al.(2005)]{2005ApJS..160..401P} Preibisch, T., Kim, Y.-C., Favata, F., et al.\ 2005, \apjs, 160, 401. doi:10.1086/432891

\bibitem[Qin et al.(2021)] {2021RAA....21...45Q} Qin, S.-M., Li, J., Chen, L., et al. \ 2021, Research in Astronomy and Astrophysics, 21, 045. doi:10.1088/1674-4527/21/2/45

\bibitem[Qin et al.(2023)] {2023ApJS..265...12Q} Qin, S., Zhong, J., Tang, T., et al. \ 2023, \apjs, 265, 12. doi:10.3847/1538-4365/acadd6

\bibitem[Rain et al.(2020)]{2020AJ....159...59R} Rain, M.~J., Carraro, G., Ahumada, J.~A., et al.\ 2020, \aj, 159, 59. doi:10.3847/1538-3881/ab5f0b

\bibitem[Rain et al.(2021)]{2021A&A...650A..67R} Rain, M.~J., Ahumada, J.~A., \& Carraro, G.\ 2021, \aap, 650, A67. doi:10.1051/0004-6361/202040072

\bibitem[Randich et al.(2013)]{2013Msngr.154...47R} Randich, S., Gilmore, G., \& Gaia-ESO Consortium\ 2013, The Messenger, 154, 47

\bibitem[Romano et al.(2021)]{2021A&A...653A..72R} Romano, D., Magrini, L., Randich, S., et al.\ 2021, \aap, 653, A72. doi:10.1051/0004-6361/202141340

\bibitem[R{\"o}ser et al.(2019)]{2019A&A...621L...2R} R{\"o}ser, S., Schilbach, E., \& Goldman, B.\ 2019, \aap, 621, L2. doi:10.1051/0004-6361/201834608

\bibitem[R{\"o}ser \& Schilbach(2019)]{2019A&A...627A...4R} R{\"o}ser, S. \& Schilbach, E.\ 2019, \aap, 627, A4. doi:10.1051/0004-6361/201935502

\bibitem[Sartoretti et al.(2018)]{2018A&A...616A...6S} Sartoretti, P., Katz, D., Cropper, M., et al.\ 2018, \aap, 616, A6. doi:10.1051/0004-6361/201832836

\bibitem[Sindhu et al.(2018)]{2018MNRAS.481..226S} Sindhu, N., Subramaniam, A., \& Radha, C.~A.\ 2018, \mnras, 481, 226. doi:10.1093/mnras/sty2283

\bibitem[Smiljanic et al.(2010)]{2010A&A...510A..50S} Smiljanic, R., Pasquini, L., Charbonnel, C., et al.\ 2010, \aap, 510, A50. doi:10.1051/0004-6361/200912957

\bibitem[Steinmetz et al.(2020)]{2020AJ....160...83S} Steinmetz, M., Guiglion, G., McMillan, P.~J., et al.\ 2020, \aj, 160, 83. doi:10.3847/1538-3881/ab9ab8

\bibitem[Sun et al.(2019)]{2019ApJ...876..113S} Sun, W., de Grijs, R., Deng, L., et al.\ 2019, \apj, 876, 113. doi:10.3847/1538-4357/ab16e4

\bibitem[Tarasov \& Malchenko(2012)]{2012AstL...38..428T} Tarasov, A.~E. \& Malchenko, S.~L.\ 2012, Astronomy Letters, 38, 428. doi:10.1134/S1063773712060059

\bibitem[Tarricq et al.(2021)]{2021A&A...647A..19T} Tarricq, Y., Soubiran, C., Casamiquela, L., et al.\ 2021, \aap, 647, A19. doi:10.1051/0004-6361/202039388

\bibitem[Wang et al.(2019)]{2019ApJS..244...27W} Wang, R., Luo, A.-L., Chen, J.-J., et al.\ 2019, \apjs, 244, 27. doi:10.3847/1538-4365/ab3cc0

\bibitem[Wang et al.(2020)]{2020ApJ...891...23W} Wang, R., Luo, A.-L., Chen, J.-J., et al.\ 2020, \apj, 891, 23. doi:10.3847/1538-4357/ab6dea

\bibitem[Zhang et al.(2020)]{2020ApJ...889...99Z} Zhang, Y., Tang, S.-Y., Chen, W.~P., et al.\ 2020, \apj, 889, 99. doi:10.3847/1538-4357/ab63d4

\bibitem[Zhong et al.(2019)]{2019A&A...624A..34Z} Zhong, J., Chen, L., Kouwenhoven, M.~B.~N., et al.\ 2019, \aap, 624, A34. doi:10.1051/0004-6361/201834334

\bibitem[Zhong et al.(2020)]{2020A&A...640A.127Z} Zhong, J., Chen, L., Wu, D., et al.\ 2020, \aap, 640, A127. doi:10.1051/0004-6361/201937131

\bibitem[Zhong et al.(2022)]{2022AJ....164...54Z} Zhong, J., Chen, L., Jiang, Y., et al.\ 2022, \aj, 164, 54. doi:10.3847/1538-3881/ac77fa



\end{thebibliography}
\end{document}